\documentclass[twocolumn,showpacs,preprintnumbers,amsmath,amssymb,floatfix,prl]{revtex4}
\usepackage{epsfig}

\begin{document}
\title{Jet production in Polarized DIS at NNLO }
\author{Ignacio Borsa}  
\email{iborsa@df.uba.ar}
\affiliation{Departamento de F\'{\i}sica and IFIBA, Facultad de Ciencias Exactas y Naturales, 
Universidad de Buenos Aires, Ciudad Universitaria, Pabell\'on\ 1 (1428) Buenos Aires, 
Argentina}
\author{Daniel de Florian}  
\email{deflo@unsam.edu.ar}
\author{Iv\'an Pedron}  
\email{ipedron@unsam.edu.ar}
\affiliation{International Center for Advanced Studies (ICAS), ICIFI and ECyT-UNSAM, 25 de Mayo y Francia, (1650) Buenos Aires, Argentina}

\begin{abstract}
We present the  NNLO calculation for single-inclusive jet production in polarized DIS $\vec{e}\vec{p} \rightarrow {\rm jet} +X$. We perform the computation based on the Projection-to-Born method by combining our recent 
 NLO result for di-jet production in polarized DIS along with the NNLO coefficients for the inclusive cross section. In this way, we achieve NNLO accuracy in a fully exclusive way for single-jet observables, the first time for a polarized cross section. We study the perturbative stability and phenomenological consequences of the QCD corrections for  Electron Ion Collider kinematics.
\end{abstract}


\maketitle


{\it Introduction.---} 
The study of the internal spin structure of the proton in terms of the contributions by quarks, anti-quarks and gluons, as codified in the (longitudinally) polarized parton distributions, is a key focus of modern particle physics.
The quark $\Delta q$ and gluon $\Delta g$ helicity distributions can be probed
in high-energy scattering processes with polarized nucleons. Experiments on fixed-target polarized 
deep inelastic lepton-nucleon scattering (DIS) performed since the late 
eighties~\cite{Aidala:2012mv} have shown that a relatively small ammount of the proton spin 
is carried by the quark and antiquark spins. 
The fixed-target inclusive DIS measurements have, however,  little
sensitivity to gluons. Instead, the best probes of $\Delta g$ were so far offered by 
polarized proton-proton collisions available at the BNL Relativistic Heavy Ion Collider 
(RHIC)~\cite{Aschenauer:2013woa}, where several processes such as 
 jet or hadron production at high transverse momentum $p_T$, receive 
substantial contributions from gluon-induced hard scattering.

Regardless of significant progress over the past three decades, and contrary to the situation in the unpolarized sector,
many open questions concerning the helicity structure of the nucleon still remain. For example, we so far have only a rather incomplete picture of the individual longitudinal polarizations of the light quarks and antiquarks, and just a first hint on the
helicity contribution of gluons inside the proton
~\cite[and references therein]{deFlorian:2009vb,deFlorian:2008mr,deFlorian:2014yva,Nocera:2014gqa,deFlorian:2019egz}.

The announcement of the approval  of the Electron Ion Collider (EIC) to be constructed at Brookhaven National Laboratory, that will collide polarized electrons and nucleons with center-of-mass energies up to $\sqrt{s}=140$ GeV, opens 
 a new era in the research on the polarized structure of the proton. Polarized DIS at larger energies allows for the study of many observables in a new kinematical range, far beyond those previous measurements at fixed-target. Among them, inclusive DIS at much smaller values of momentum fraction $x$ and larger virtuality $Q^2$ and jet production would grant access to definite knowledge on the polarized gluon content of the proton. 
Clearly, the extraction of information on the spin structure of the nucleon requires both accurate measurements and precise theoretical evaluations for the corresponding observables. In perturbative QCD, cross-sections are computed as an expansion in the strong coupling constant $\alpha_s$. 
Leading order (LO) calculations ${\cal O}(\alpha_s^0)$ usually present only a {\it qualitative} description of an observable since higher order corrections are known to be large and needed to provide reliable {\it quantitative} predictions for a high-energy process. 
Besides reaching a higher accuracy, one of the key issues is to check the perturbative stability of the process considered, that is, to examine to which extent  the higher order corrections affect the cross sections and the  asymmetries relevant for experimental measurements. Only if the corrections are under control can a process that shows good sensitivity to a given  parton density be considered as a genuine probe for that, and be reliably used to extract accurate distributions from future data. 

Furthermore, the inclusion of extra partons in  higher order perturbative calculations is particularly important in the case of jet production, since it is only from next-to-leading order (NLO) that the QCD structure of the jet starts to play a role in the theoretical description, providing the possibility to realistically match the experimental conditions imposed to define a jet. 
In general, a better  description of the jet can be achieved when more partons are included in the final state.
This is particularly obvious when the calculation is performed at the fully exclusive level, such that the four-momenta of all outgoing particles (leptons and partons) become available in order to apply the same cuts used at the experimental level. 

In the case on unpolarized colliders, such as the Large Hadronic Collider (LHC), next-to-next-to leading order (NNLO) computations (i.e.${\cal O}(\alpha_s^2)$ with respect to the corresponding lowest order) have become the state-of-the-art, and some observables have even already reached the next level. A similar situation occurs for unpolarized DIS, where both the inclusive \cite{Vermaseren:2005qc} and  single-inclusive jet \cite{Gehrmann:2018odt} cross sections have been computed up to next-to-next-to-next-to leading order (N$^3$LO), see e.g.\cite{Amoroso:2020lgh}.

The situation is quite different in the polarized sector. While a number of observables were computed to NLO accuracy, only a couple of fully inclusive cross sections, such as DIS~\cite{Zijlstra:1993sh} and Drell-Yan \cite{Ravindran:2003gi}, as well as the helicity-dependent splitting functions \cite{Vogt:2008yw,Moch:2014sna,Moch:2015usa}, are known to the next order, NNLO. Clearly, the advent of the EIC requires to level up the state-of-the-art in polarized cross sections to the ones reached for the corresponding unpolarized counterparts in order to perform a detailed study of asymmetries. 
In particular, jet production in polarized lepton-nucleon scattering has been computed to NLO accuracy in the photoproduction domain either analytically in the {\it small-cone approximation} \cite{Hinderer:2015hra,Jager:2008qm} or as a fully exclusive Monte-Carlo implementation \cite{deFlorian:1999ge}.  A recent calculation for single-inclusive jet production \cite{Boughezal:2018azh}, based on the polarized extension of the \textit{N-jettiness} subtraction scheme \cite{Boughezal:2017tdd}, accounts for  DIS at NLO and for the lowest order contribution in the photoproduction regime.

In this letter we present the first NNLO calculation for single-inclusive jet  production in polarized DIS $\vec{e}\vec{p} \rightarrow 1\,{\rm jet}$. This is achieved by combining our recent fully exclusive computation for di-jet production at NLO in polarized $\vec{e}\vec{p} \rightarrow 2\,{\rm jets}$ collisions \cite{inpreparation} with the NNLO expression for inclusive DIS~\cite{Zijlstra:1993sh} by means of the 
Projection-to-Born  (P2B) method \cite{Cacciari:2015jma}. We study the perturbative stability and phenomenological effects on both polarized and unpolarized cross sections, with the corresponding asymmetries, in terms of the most relevant variables for the process
at EIC kinematics.

{\it NNLO corrections to DIS.---} 
We specify the process $e(k)+P(P) \rightarrow e(k') + {\rm jet}(p_T,\eta) +X$ where $k$ and $P$ are the momenta of the incoming electron and proton, respectively and $k'$ is the momentum of the outgoing (detected) electron. We consider only photon exchange such that its four-momentum is given by $q=k-k'$ and its virtuality by $Q^2=-q^2$. The usual electron and Bjorken variables are given by, $y=P\cdot q/P\cdot k$ and $x=Q^2/(2 P\cdot q)$ respectively. The final state jet is characterized by the transverse momentum $p_T$ and pseudorapidty $\eta$. We analyze the process in the laboratory frame of the lepton–proton system (LAB) where the jet has always non-vanishing transverse momentum, with a contribution starting already at ${\cal O}(\alpha_s^0)$. This is at variance with the Breit-frame, where the proton and the virtual photon collide head-on and the LO term does not contribute to the production of jets, starting only from di-jet production at ${\cal O}(\alpha_s)$.

The computation of higher order QCD corrections to any process is complicated due to the appearance on many infrared singularities that cancel when adding all real, virtual and factorization contributions. There are several methods to handle the singularities in the intermediate steps of the calculation, either {\it subtraction} or {\it slicing} based. One of the simplest is the Projection-to-Born method \cite{Cacciari:2015jma}, which defines the subtraction by the full matrix element evaluated at the original phase space point but binned in the kinematic corresponding  to the Born-projected equivalent for the lowest order process. The method is restricted only to processes such as the production of colourless particles in hadronic collisions or single-inclusive jet  production in DIS in the LAB frame, where the  Born-projected kinematics can be reconstructed from the momenta of non-QCD particles. For those observables ${\cal O}$, one can write the cross section at  ${\rm N^kLO}$ accuracy as
\begin{equation}
d\sigma_{{\cal O}}^{ \rm N^kLO} = d\sigma_{{\cal O}+jet}^{\rm N^{k-1}LO} -d\sigma_{\rm {\cal O}+jet\, P2B}^{\rm N^{k-1}LO} + 
d\sigma_{\rm {\cal O}\, P2B}^{\rm N^{k}LO , incl}
\end{equation}
where the first term represents the result for the same observable (single-jet production  in this case)  plus one extra jet at  ${\rm N^{k-1}LO}$ accuracy, the second is the subtraction term corresponding to the same quantity as before but now {\it binned} at the P2B kinematics, while the last one corresponds to the fully inclusive result at the same desired accuracy.
In the case of DIS, the P2B kinematics is simply constructed by noting that the lowest order partonic process is characterized by  $e(k)+q(p) \rightarrow e(k') + q(p')$. The P2B algorithm is then defined by keeping the original momenta of the event for the electrons ($k$ and $k'$) and  mapping the momenta of the incoming and outgoing partons by the Born relations $p=x P$ and $p'=x P+q$.
Therefore, in order to reach NNLO accuracy for the process of interest one needs an exclusive calculation for {\it di-jet} production in polarized DIS at NLO and the NNLO expression for the polarized structure function, as presented in  \cite{Zijlstra:1993sh}. For the first ingredient, we have performed the corresponding di-jet NLO computation using a modified version of the {\it dipole subtraction method} \cite{Catani:1996vz} that accounts for spin dependent effects \cite{inpreparation}. The calculation is implemented in the code {\tt POLDIS} that allows to compute any infrared safe observable related to single-jet production at NNLO accuracy, as well as to single- and di-jet production in the Breit-frame with NLO precision \footnote{The code is partially based on DISENT, written by M, Seymour \cite{Catani:1996vz}. A bug found in one of the dipole terms for the gluon channel was corrected. This correction solves the disagreement with DISASTER reported in \cite{Antonelli:1999kx},\cite{Dasgupta:2002dc}, \cite{McCance:1999jh} and \cite{Nagy:2001xb}}. Here we concentrate on single-jet observables, further phenomenological results involving more jets will be shown elsewhere \cite{inpreparation}.

In order to present the NNLO results for the EIC, considering the configuration with a proton beam energy $E_p =275$ GeV and electron beam energy  $E_e =18$ GeV, we rely on the following set-up. For unpolarized and polarized parton distributions we use the NLO PDF4LHC15 \cite{Butterworth:2015oua} and DSSV \cite{deFlorian:2014yva,deFlorian:2019zkl} sets, respectively, and fix the central factorization and renormalization scales to  $\mu_F^2=\mu_R^2=Q^2$ with $\alpha_s$ evaluated also at NLO accuracy with $\alpha_s(M_Z)=0.118$.
Jets are reconstructed using the anti$k_T$ clustering algorithm with $R=0.8$ using the $E_T$-weighted recombination scheme, and are required to satisfy 
\begin{equation}
5\, {\rm GeV} < p_T <  36\, {\rm GeV} \,\,\,\, {\rm and} \,\, |\eta|<3 \, .
\end{equation}
Furthermore, on the leptonic side we request cuts similar to those of HERA, with
\begin{equation}
25\, {\rm GeV}^2 < Q^2 <  1000\, {\rm GeV}^2 \,\,\,\, {\rm and} \,\, 0.04<y<0.95 \, .
\end{equation}
Here, the lower limit in $Q^2$ is set considering that at LO the transverse momentum of the jet is given by $p_T^{2\,\,(LO)}=Q^2(1-y)$. Notice that for $Q^2\lesssim 25$ GeV the calculation is actually one-order less accurate since the Born-level contribution is kinematically not allowed.
\onecolumngrid
~

\begin{figure}[h]
 \epsfig{figure=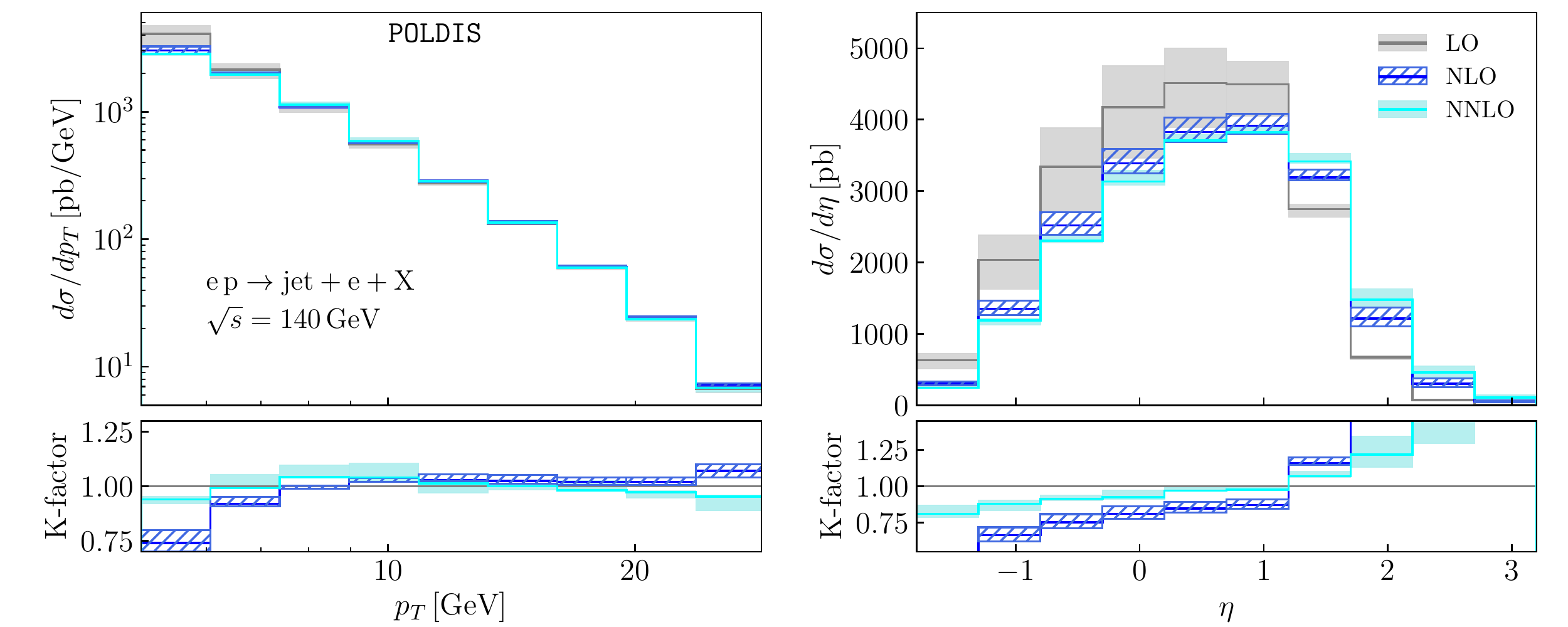,width=0.99\textwidth}
  \caption{Single-jet pseudorapity and transverse momentum unpolarized distributions at LO, NLO and NNLO. The bands  reflect the variation in the cross-section when changing the scales as $\mu_R=\mu_F=[1/2,2] Q$.
  The lower inset shows the corresponding $K-$factors as defined in the main text.}\label{fig1}
\end{figure}
 \twocolumngrid

\vspace{0.2cm}
\onecolumngrid
~
\begin{figure}[h]
 \epsfig{figure=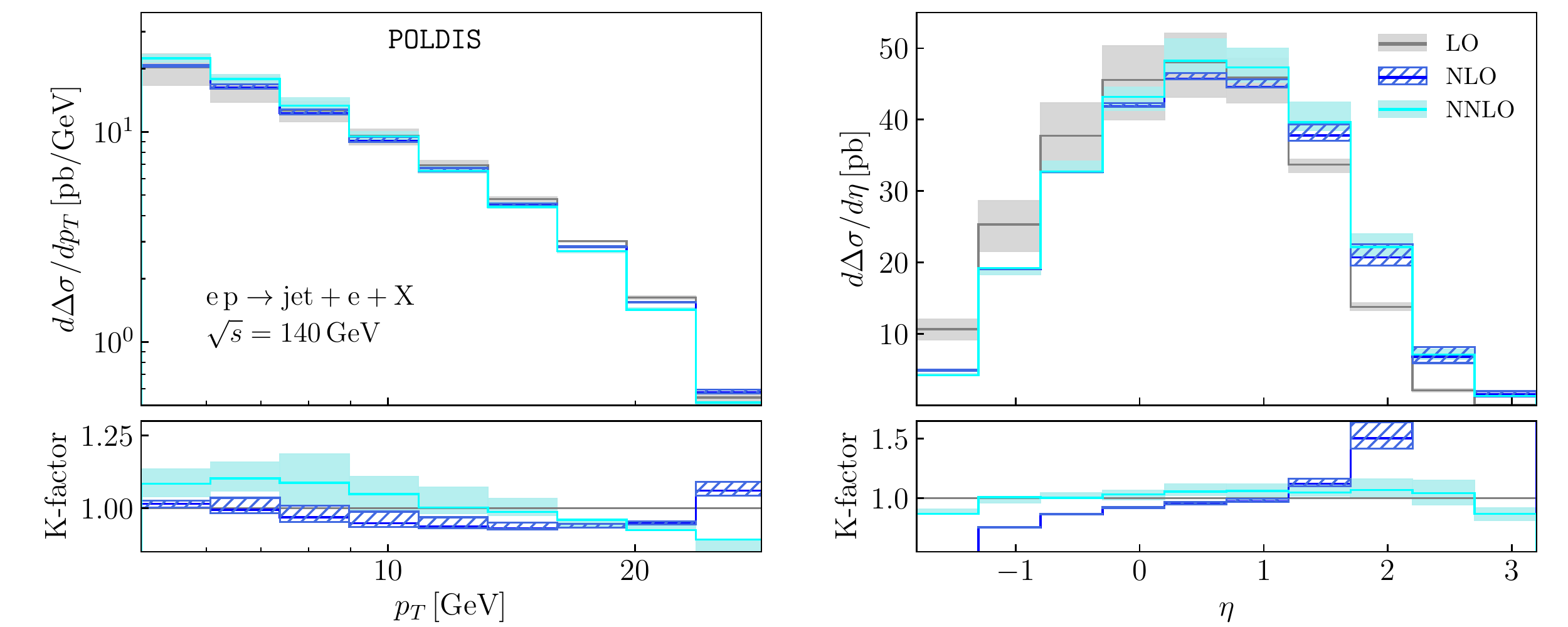,width=0.99\textwidth}
 \caption{Same as Fig.\ref{fig1} but for  {\it polarized} deep inelastic scattering.}\label{fig2}
 \end{figure}
 \twocolumngrid

In Figure \ref{fig1} we present the unpolarized cross section for single-inclusive jet production calculated at LO, NLO and NNLO accuracy in terms of the pseudorapidity $\eta$ and the transverse momentum $p_T$ of the jets. The lower inset shows the corresponding $K-$factors defined as $K^{\rm NLO}=\sigma^{\rm NLO}/\sigma^{\rm LO}$ and $K^{\rm NNLO}=\sigma^{\rm NNLO}/\sigma^{\rm NLO}$ in order to quantify the effect of the  corrections at each subsequent order.

As it happens in the case of HERA \cite{Gehrmann:2018odt},  a clear trend can be observed on the effect of higher order corrections on  pseudorapidity, with rather small corrections in the central region but larger contributions in the forward region $\eta>1$, which becomes populated by the extra jets generated  at NLO and NNLO. In that kinematical regime, higher order corrections become essential for an accurate description of the distribution. The {\it shift} in the distribution towards larger rapidities also results in a considerable reduction in the cross-section in the electron beam direction ($\eta<-1$).
In the same way, the transverse momentum distribution is also much affected by higher-order corrections in the 
low $p_T$ region. In Figure \ref{fig1} we also present a first estimate of the perturbative uncertainties by performing simultaneous variations of the renormalization and factorization scales as $\mu_R=\mu_F=[1/2,2] Q$, represented by the corresponding  bands. In general we observe a reduction in the size of the band when moving to higher orders, and rather good overlap between the NLO and NNLO bands, anticipating an improvement in the convergence of the expansion.

Figure \ref{fig2}  introduces the same observables studied in  Fig.\ref{fig1} but in the polarized case. The pattern for the corrections are roughly the same  but some differences already arise. For example, the scale dependent bands at NLO are smaller than the ones in  unpolarized collisions, and for some bins even smaller than the NNLO ones. This is due to the particular behaviour of the polarized parton distributions and partonic cross sections, such that the  NLO polarized structure function $g_1(x,Q^2)$ presents a change of sign in the relevant   $(x,Q^2)$ domain for these observables. We also observe an overlap between the NLO and NNLO bands for all bins in both pseudorapidity and transverse momentum distributions. Nevertheless, the scale variation presented here should be considered only a a first attempt to quantify the perturbative stability but not as true estimate of the size of the missing higher orders \footnote{Furthermore, notice that due to the lack of NNLO polarized pdf sets all cross sections were evaluated with NLO evolved distributions without being able to match the NNLO accuracy from the non-perturbative side}.

\begin{figure}[h]
  \epsfig{figure=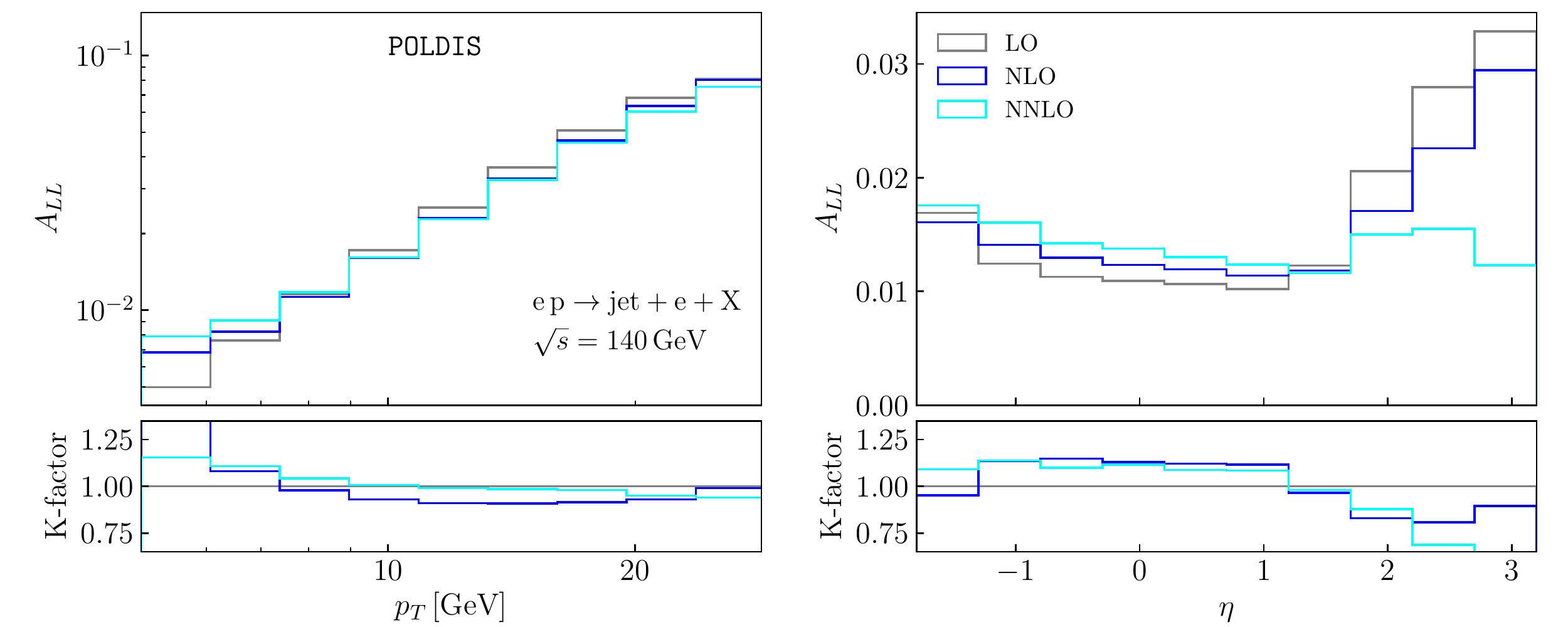,width=0.95\textwidth}
  \caption{Double spin asymmetries $A_{LL}$ at  LO, NLO and NNLO. }\label{fig3}
\end{figure}

Finally, we look at the most relevant observables in polarized collisions, the double spin asymmetries, defined as the ratio between the corresponding polarized and unpolarized cross sections $A_{LL}=\Delta \sigma / \sigma$. The  LO, NLO and NNLO results for EIC kinematics, with asymmetries at the level of $A_{LL}\sim{\cal{O}}(1-2\%)$ are presented  Figure \ref{fig3}.

Even though polarized and unpolarized distributions show similar features, the effect of the higher order corrections in the asymmetries is not negligible, as can be observed in the $K-$factors presented in Fig.\ref{fig3}, which shows a NNLO correction that can reach about $15\%$ with respect to NLO. In this sense, it is worth noticing that the NNLO enhancement of the unpolarized cross-section in the forward region leads to a sizable suppression of the double spin asymmetry for $\eta>1.5$. At the extreme kinematical regime of  very forward jet production $\eta>2.5$, and for the particular set of cuts applied (especially on $y>0.04$), the fixed order perturbative expansion  does not show convergence and the resummation of large logarithmic corrections would become necessary. 
A somewhat milder enhancement is obtained for the central region in pseudorapidity and higher order corrections turn out to be also relevant for an accurate description of $A_{LL}$ for low values of $p_{T}$, for which a rise in the asymmetry is observed. 

In terms of the share between initial state partons, we find that the asymmetry for single-jet production is mostly determined by the quark (and antiquark) polarization, given the relevance of that channel  open at the Born-level. The polarized gluon accounts for a negative contribution of the order of $-5\%$ to $-10\%$ of the total asymmetry at central pseudorapidities, as expected since it starts only from NLO. Stronger constraints on the polarized content of the gluon could be obtained from di-jet production in the Breit-frame, with an  initial state gluon channel open already at its lowest order $\alpha_s$.
The uncertainties arising from the polarized parton distributions turn out to be of the order of  $5\%$ to $10\%$ over the kinematical range studied here, comparable and for some bins even smaller than the size of the higher order corrections. It is clear, therefore, that for a more precise extraction of polarized parton distributions from those observables at the EIC it is essential to consider the perturbative QCD expansion up to NNLO accuracy.

{\it Conclusions and outlook.---} 
 In this letter we have presented the first NNLO fully-exclusive calculation for single-inclusive jet production in polarized DIS. This was archived by the projection-to-Born method, which utilises our computation of fully exclusive NLO polarized di-jet production in $\vec{e}\vec{p}$ collisions along with the inclusive NNLO polarized structure function. The calculation was implemented in our new code \texttt{POLDIS} which allows to compute any infrared-safe observable in single- and di-jet production up to NNLO and NLO accuracy, respectively. 
 
 We presented the single-inclusive jet production results in the kinematics of the future EIC, in terms of the jet pseudorapidity $\eta$ and its transverse momentum $p_T$. The differential distributions show sizable corrections at NNLO in both the polarized and unpolarized cases, that shifts the rapidity distribution towards the forward region, as observed previously in HERA kinematics \cite{Gehrmann:2018odt}. In both cases, we found good agreement between the NLO and NNLO corrections, with overlapping theoretical uncertainties, indicating convergence of the perturbative expansion. The impact of higher order corrections was also observed in the corresponding double spin asymmetries, which are suppressed at NNLO in the forward region, albeit with larger theoretical uncertainties. 
 
 Our results highlight the importance that higher order corrections in $\alpha_{s}$ will have in a precise description of the observables to be measured in the EIC and ultimately in the improvement of our picture of the spin structure of nucleons.

\acknowledgments
We thank  R. Sassot, W. Vogelsang and M. Stratmann for discussions and useful communications.
This work was partially supported by CONICET and ANPCyT.      
\onecolumngrid
~
\twocolumngrid

\bibliography{paper}

\end{document}